# Long Thermal Stability of Inverted Perovskite Photovoltaics Incorporating Fullerene-based Diffusion Blocking Layer

*Fedros Galatopoulos [1]Ioannis T. Papadas [1], Gerasimos S. Armatas[2] and Stelios A. Choulis [1]\**

[1] Molecular Electronics and Photonics Research Unit, Department of Mechanical Engineering and Materials Science and Engineering, Cyprus University of Technology, Limassol, 3603 (Cyprus).

[2] Department of Materials Science and Technology, University of Crete, Heraklion 71003, Greece.

*Email: stelios.choulis@cut.ac.cy

In this article, the stability of p-i-n perovskite solar cells is studied under accelerated heat lifetime conditions (60 $^o$C ,85$^o$C and $N_2$ atmosphere). By using a combination of buffer layer engineering, impedance spectroscopy and other characterization techniques, we propose the interaction of the perovskite active layer with the top Al metal electrode through diffusion mechanisms as the major thermal degradation pathway for planar inverted perovskite photovoltaics (PVs) under 85$^o$C heat conditions. We show that by using thick solution processed fullerene buffer layer the perovskite active layer can be isolated from the top metal electrode and improve the lifetime performance of the inverted perovskite photovoltaics at 85 $^o$C. Finally, we present an optimized solution processed inverted perovskite PV device using thick fullerene-based diffusion blocking layer with over 1000 hours accelerated heat lifetime performance at 60ºC.

## 1. Introduction

Perovskite-based solar cells have been amongst the most prominent photovoltaic (PV) technologies over the last couple of years offering a tremendous advancement in power conversion efficiency (PCE) from 3.8% to 22.7%.[1,2] Several characteristics such as high absorption coefficient,[3] tunable bandgap,[4,5] long carrier diffusion length,[6] high carrier mobility,[7] and low exciton binding energy,[8,9] have allowed perovskite-based solar cells to reach such high values in PCE, even if casted from solution at low processing temperatures.[10]

Although important advancements have been made to improve PCE, the long-term stability of perovskite-based devices still remains a challenge however. Among others, important factors that affect the stability of perovskite solar cells are the moisture, heat and light irradiation. These factors can deteriorate the performance of both the active layer and electrodes of the device. The organic-inorganic perovskite semiconductor materials are known to be intrinsically unstable to moisture [11,12] since they can form hydrated crystal phases, which mobilize the organic species, and decompose the crystal to lead iodide ($PbI_2$).[13] Decomposition of the perovskite active layer to $PbI_2$ can also happen in the presence of $O_2$ and light. It was suggested that superoxide ions ($O_2^-$) can be formed upon irradiation of $CH_3NH_3X_3$ (X=Cl, Br, I) halides which, in turn, deprotonate the methylammonium cation ($MA^+$) and decompose the perovskite structure. [13, 14]

As a result, considerable research efforts have been dedicated towards the improvement of light and especially humidity stability of perovskite based solar cells. It has been previously shown that under accelerated humidity conditions, there is a significant active layer degradation due to the decomposition of the perovskite crystal. Shielding the active layer from humidity, and therefore, improvement in device lifetime has been reported through the usage of hydrophobic layers such as thiols.[11] Fan et. al have also reported efficient moisture resistance by incorporating a cerium oxide (CeOx) bilayer achieving stability of 200 h under ambient conditions. [15] In a recent work, Yang Z. et. al. have reported that the stability of perovskite solar cells under continues illumination can also be improved by doping the [6,6]-Phenyl-butyric acid methyl ester (PCBM) with graphene quantum dots (GQDs).[16] More recently, it has been reported that photo-oxidation of the perovskite active layer leads to the formation of charge trap centers which, in turn, may cause significant changes in charge density and charge extraction efficiency of the device. [17] On the other hand, by incorporating mixed-cation based devices, such as formamidinium-cesium ($FA_{0.83}Cs_{0.17}Pb(I_{0.8}Br_{0.2})$), the oxygen-induced degradation can be markedly minimized.[18] Research regarding the heat stability of such devices has been scarce however, especially at elevated temperatures such as 85 °C. The 85°C is a particularly interesting temperature since, according to International Standards (IEC 61646 climatic chamber tests), it represents the elevated temperature of a roof during a hot summer day. [19] Although $CH_3NH_3PbI_3$ has been previously reported to be mostly intrinsically stable at elevated temperatures as high as 85 °C in the absence of $O_2$ and ambient conditions, [19, 20] the devices themselves are largely unstable. Zhao X. et. al have reported that spiro-MeOTAD exhibits a state change at elevated temperatures, which reduces the hole mobility. These issues can be tackled by using the thermally stable NiO and PCBM materials as the hole transporting layer (HTL) and electron transporting layer (ETL), respectively.[20] In another work where the active layer was damaged by subsequent electrode interaction was reported by Domanski K. et. al. In their work they suggested that temperatures as low as 70 ºC can cause Au metal migration towards the active layer. The migration of Au can be blocked by incorporating a Cr interlayer between the Au metal electrode and HTL layer. [21] Using recent modelling tools it has been also proposed that blocking layers can reduce the voltage drop on perovskite active layer, providing reduced ionic motion and predicted improved stability. [22] In a more recent work, Bi et. al have reported improved heat stability by using an ETL consisting of N-doped graphene, PCBM and CQDs. [23]

In this work we have performed accelerated heat lifetime tests at 85 ºC under inert ($N_2$) atmosphere in the dark, using p-i-n perovskite-based solar cells with the following structure: ITO/PEDOT:PSS/$CH_3NH_3PbI_3$/PC[70]BM/AZO/Al, following the perovskite formulation reported by Liang et. al. [24] [6,6]-Phenyl-C70-butyric acid methyl ester (PC[70]BM) is a widely used material in p-i-n perovskite solar cells that not only functions as an ETL but it can also passivate charge trap sites present on the surface and grain boundaries of the perovskite films. [25] Aluminum-doped zinc oxide (AZO) was also incorporated since it has been previously reported to increase the electron selectivity of such device structures. [4] Thus, both PC[70]BM and AZO ETLs play an essential role on the device operation of p-i-n perovskite PVs. In this report two different thicknesses of PC[70]BM were used for heat lifetime performance tests. These include $CH_3NH_3PbI_3$-based devices with PC[70]BM layers of ~70 nm (thin fullerene-based devices) and ~200 nm (thick fullerene-based devices) thickness. We show that thick PC[70]BM yields improved heat stability over the thin PC[70]BM counterparts at the expense of reduced PCE. We have shown that PC[70]BM layer isolation is the key for achieving improved stability. Finally, we also show a perovskite formulation based on $CH_3NH_3PbI_{3-x}Cl_x$ that is optimized to function efficiently using a thick PC[70]BM electron transporting layer (ETL) while still maintaining good PCE. This developed device retained very good thermal stability at 60 ºC for 1000 h.

## 2. Results and Discussion

### 2.1   CH$_3$NH$_3$PbI$_3$ film characterization

CH$_3$NH$_3$PbI$_3$ (MAPbI$_3$) has been previously reported to be mostly intrinsically stable at 85 °C under N$_2$ atmosphere by Conings et. al. [19] In order to confirm previous observations, ITO/PEDOT:PSS/CH$_3$NH$_3$PbI$_3$ films were annealed at 85 °C on a hotplate inside a N$_2$ filled glovebox in the dark for 96 hours. The XRD patterns of fresh and thermally annealed films are shown in **Figure 1**. Similar to what has been previously reported in the literature, [17] the XRD patterns show no traces of PbI$_2$, judging by the absence of diffraction peaks at 12.7° and 38.7°. Although there is a peak at 52.4° which has been previously attributed to PbI$_2$ phase, the size of this peak does not change upon thermal treatment. This suggest that CH$_3$NH$_3$PbI$_3$ is, for the most part, intrinsically stable at temperatures to 85 °C under N$_2$, which is in agreement with previous studies.[19] The UV-Vis absorption spectra of fresh and aged perovskite films also remain quite unchanged and are similar to the ones reported by Conings et.al which further points towards the intrinsic stability of the perovskite film. (**Figure S1**)

### 2.2. Heat stability and characterization of CH$_3$NH$_3$PbI$_3$ based devices

As it was previously mentioned, we tested two different sets of p-i-n solar cells incorporating different fullerene thicknesses, i.e. PC[70]BM with ~70 nm (thin fullerene-based devices) and PC[70]BM~200 nm (thick fullerene-based devices) film thickness. The devices were encapsulated using a UV-curable encapsulation epoxy and a small glass slide in order to avoid any possible ingress of moisture and then placed on a hotplate at 85 °C in a $N_2$-filled glovebox under dark. The devices were taken outside of the glovebox to measure the PV device performance parameters and undergo the characterization process. The devices were then placed back again inside the glovebox to continue the aging process. This procedure was repeated at 24h intervals for up to 168 h. The deterioration of the normalized photovoltaic parameters was monitored over the course of 168 hours at 24-hour intervals for both device sets, and is shown in **Figure 2**.

From Figure 2 it can be seen that there are striking differences between the two sets of devices. The thin-fullerene based device exhibits a very sharp drop to all of their photovoltaic parameters only just after 24 hours of heating. The drop in photovoltaic parameters seems to stabilize at the 72-h mark. However, the device PCE has already dropped well below 20% by that point. On the other hand, thick-fullerene based device retains almost 100% of the PCE even after 168 hours of heating. Although there is a noticeable drop of the Voc at ~80% at the first 24 hours, there is an equally increase in Jsc (115%), and especially FF (140%), for this device. Both the drop in Voc and increase in Jsc and FF are stabilized after 24 hours, allowing the thick-fullerene based device to preserve its high PCE (~100 %) even after 168 hours of heating.

Representative devices from each device configuration were chosen for characterization. The representative devices yielded the following photovoltaic parameters: thin-fullerene based device (Voc=0.91 V, Jsc=14.75 mA/cm$^2$, FF=79.2 %, PCE=10.64 %), thick-fullerene based device (Voc=0.87 V, Jsc=10.57 mA/cm$^2$, FF=40.8 %, PCE=3.75 %). These measurements indicated a considerable decrease in FF when we increase the thickness of PC[70]BM, due to the limited electrical conductivity of the material. [4] The purpose of this study was to identify the major degradation pathway of the $CH_3NH_3PbI_3$ formulation-based devices under accelerated heat conditions and to evaluate the effect of different fullerene thickness on the device lifetime performance. For the above reasons achieving the highest possible PCE for each device set was not within the targets of this experimental plan.

**Figure 3** shows the illuminated J/V characteristic of fresh and aged devices after 96 hours of heating. The 96 hours mark was chosen as the point of degradation where all the characterization studies were performed on the examined samples. The reason for this choice was that at the 96 hours mark the decline of the photovoltaic parameters has been substantially reduced. As it can be seen from Figure 3, the shape of the J/V curve for thin-fullerene based devices is severely distorted after 96 hours of heating, which is consistent with the severe drop on photovoltaic parameters, as evident from Figure 2. Furthermore, significant signs of hysteresis start to appear within the device performance characteristics, which were absent from the fresh devices. Hysteresis in perovskite solar cells has been previously attributed, amongst other phenomena, to non-steady state capacitive currents resulting from electrode polarization. [26] Thus, the manifestation of hysteresis in the J/V plot could point to electrode degradation in the device, which prevents the efficient extraction of carriers. On the other hand, the thick-fullerene based devices don't exhibit any shape deterioration of the J/V plot or any hysteresis behavior after degradation. This correlates well with the stable performance of the thick-fullerene based devices, where the initial values of the photovoltaic parameters are mostly retained. Furthermore, the J/V plots of thick-fullerene based devices show an S-shaped profile growth prior to the heating test, which completely disappears after 96 hours of annealing. An S-shape in the J/V curves has been previously attributed to poor charge extraction, which in our devices can be related with the large thickness of the PC[70]BM layer. [27] The absence of the S-shape after 96 hours heating could be a result of the light soaking effect from the constant light illumination during the characterization of the devices. It has been previously reported that this phenomenon is a result of the trap-assisted recombination of excitons[28], which can be strengthened due to the increase in trap density in the thick PC[70]BM film. The different Jsc behavior from both the lifetime plots and illuminated JV characteristics was also observed when we performed photocurrent (PCT) measurements (**Figure S2**). PCT is a very useful technique for degradation studies that has been previously used in organic photovoltaics (OPVs) as well [29] for the visualization of how the current intensity is distributed inside a device. As it was expected, the device incorporating thin PC[70]BM layer exhibited a severe loss to current intensity, whereas the device with thick PC[70]BM layer retained its current intensity to a good degree.

Complementary to the previous characterization techniques, impedance spectroscopy was also used. Impedance spectroscopy is a very powerful technique that can be used to provide insight on the physical processes occurring inside the device. [3] The impedance spectroscopy data were analyzed and represented using Nyquist as well as Mott-Schottky plots.

**Figure 4a** shows the standard shape of two frequency responses for perovskite solar cells (a high- and low-frequency feature). The feature at high frequencies has been previously attributing to charge transport resistance (Rtr) of the hole transporting layers (HTLs) and electron transporting layers (ETLs) as well as their interface with the perovskite active layer.[30, 31] The low frequency feature has been attributed to the recombination resistance (Rrec) and ionic diffusion.[30-32] From the results in Figure 4a, a significant decrease in Rrec is observed between fresh and aged thin-fullerene based devices as well as an increase in Rtr. The increase to Rtr denotes that the ETL/HTL or their interface with the perovskite has been altered, making carrier movement inside the device more difficult. This in return results to a decrease of Rrec due to more frequent charge recombination events. The combination of increasing Rtr and decreasing Rrec is in good agreement with the decreased FF (~40%) noted for thin-fullerene based devices, highlighting the difficulty for carrier extraction. On the other hand, the Rtr for thick-based devices remain relatively unchanged, whereas a slight increase of the Rrec was observed which could be tied with the normalized FF increase.

Mott-Schottky analysis is often used for the differentiation of processes occurring at the active layer with the ones occurring at the interfaces and outer contacts of the device. [33] The capacitive plateau at negative voltages of **Figure 4b** provides information on the dielectric constant, which is an intrinsic property of the bulk material. A decrease in Vbi is observed (point of intersection of the slope with the x-axis) for both thin and thick fullerene-based devices, which denotes that the energy equilibration at the contacts has shifted. [33] More importantly, Vbi for thin-ETLs based devices has dropped by ~0.5 V whereas for thick-fullerene based devices only by ~0.25 V. This is in accordance with the more rapid decrease of normalized Voc for the thin-ETLs based devices compared to the thick-fullerene based devices, where the Voc drop is a lot less apparent and more stable.

### 2.3. Buffer layer device engineering experimental methods

In order to try and isolate the effect of heat in each layer of the device, buffer layer engineering methods were used. This is a powerful technique, which was also previously used in organic photovoltaics (OPVs) [29] allowing  us to study the effect of the PV parameters in a fabricated device using aged films at 85oC for 96 hours, similar to the conditions used for device characterization. To achieve our goal, semi-finished device structures were initially fabricated using aged ITO/PEDOT:PSS/CH$_3$NH$_3$PbI$_3$ and aged ITO/PEDOT:PSS/CH$_3$NH$_3$PbI$_3$/PC[70]BM/AZO device structures at 85$^o$C for 96 hours. The photovoltaic parameters of the aged device structures completed with fresh PC[70]BM/AZO/Al and Al respectively were compared with the values of fresh (not-degraded) reference devices. The results are shown in **Table 1.**

From Table 1 aged device structures (ITO/PEDOT:PSS/CH$_3$NH$_3$PbI$_3$ and aged ITO/PEDOT:PSS/CH$_3$NH$_3$PbI$_3$/PC[70]BM/AZO at 85 $^o$C for 96 hours with fresh PC[70]BM/AZO/Al and fresh Al retain their photovoltaic performance with minor losses compared to the fresh reference devices. This further highlights the intrinsic stability of MAPbI$_3$ and the important role that the top electrode interaction with the perovskite active layer plays for the stability of the device. If we compare the device structures which were fabricated using aged MAPbI$_3$ with the aged devices of FIG. 3, aging the MAPbI$_3$ doesn't play a significant role to the degradation of the device since at 96 hours the PCE drops below ~20% of its initial value, whereas the device structures (with aged active layer and fresh PC[70]BM/AZO/Al) retained most of their initial PCE value. A similar behavior is observed when fresh Al was evaporated within an aged ITO/PEDOT:PSS/CH$_3$NH$_3$PbI$_3$/PC[70]BM/AZO device structure. The resulting devices, like before, exhibit a very similar behavior comparable to our non-degraded fresh reference devices. From the above experimental buffer layer device engineering observations, we conclude that the devices degrade only upon the incorporation of Al during the thermal aging test and therefore the interaction of the perovskite active layer with the top metal electrode is the major degradation pathway for this type of device, whereas the perovskite itself or its interaction with any subsequent interlayers plays no significant role in the heat stability of the devices.

Two possible mechanisms are proposed that could be related to the degradation of the devices: i) migration of halide ions to the Al metal and ii) migration of Al atoms to the MAPbI3. Both these mechanisms were also observed by Fang et. al where migration of I⁻ towards the Ag electrode as well as migration of Ag atoms into the MAPbI3 manifested under ambient humidity and light soaking for 200h. [15] In particular, diffusion of I⁻ is common in such devices due to the small activation energy (~0.1 eV) [34] and high concentration gradient ($10^{26}$ cm$^{-4}$) which is further accelerated at higher temperatures. [23] Diffusion of I⁻ is often facilitated by the decomposition of the perovskite layer. During decomposition of the perovskite layer to lead iodide (PbI2), methyl ammonium (MA) and hydrogen iodide (HI), MA and HI can escape the surface of the perovskite leaving behind iodide vacancies which promote the diffusion of I⁻ [23]. Metal ions have also been reported to migrate towards the perovskite under accelerated heat conditions. Domanski et. al have recently reported that upon heating at 70 ºC, Au atoms can travel through spiro-MeOTAD HTL and migrate towards the perovskite layer, remarkably affecting the device performance. [20] The direct contact of the perovskite layer with metal electrodes like Ag has also been previously reported to be very damaging for the device performance even at ambient conditions [23, 35] due to the chemical interaction between the two materials. Therefore the isolation of these two layers is of utmost importance.  Al and other commonly used electrode metals (Au, Ag and Ca) have been previously reported to penetrate into fullerene layers even in the process of evaporation. This increases the probability of such metal and the perovskite layer to come in direct contact with each other chemically reacting.[35, 36] This chemical reaction between the metal electrodes and perovskite leads to a change to the perovskite film color from dark brown to yellow due to the decomposition to PbI$_2$. [23] **Figure 5** shows devices presented within this paper: a) bottom side of fresh device stack, b) bottom side of thermally aged device stack of thin-PC[70]BM based devices, c) bottom side of thermally aged devices stack of thick-PC[70]BM based devices, d) the top side of fresh device stack, e) top side of thermally aged device stack of thin-PC[70]BM based devices, f) top side of thermally aged devices stack of thick-PC[70]BM based devices .

From Figure 5b we can see that the perovskite film in thin-PC[70]BM based devices exhibited a color change from dark brown to yellow upon thermally aging the whole device stack at 85ºC and N$_2$ similar to previously reported observations. [23]. In the contrary the perovskite retains its color at a respectable degree for devices incorporating thick PC[70]BM (Figure 5c). Furthermore metallic-like spots start to appear in the thin-PC[70]BM based devices located at the bottom side of the device which were absent from both the fresh and thick-PC[70]BM based devices. The top side of the thin PC[70]BM device also shows some change in the color of Al electrode where it became less shiny and more white (Figure 5e). This is an indication of Al corrosion due to diffusion of I$^-$ as previously reported.[23] Both the fresh (Figure 5d) and aged thick-fullerene based devices (Figure 5f) don't show any apparent change in the Al electrode.

Recently reported time of flight secondary ion mass spectroscopy (ToF-SIMS) measurements have shown metal diffusion of air stable metal such as Au in a Cs containing FA$_{0.83}$MA$_{0.17}$Pb(I$_{0.83}$Br$_{0.17}$)$_3$ formulation while in contrast, the ToF-SIMS profile of I$^-$ remained relatively unchanged between fresh and thermally aged devices, pointing to the limited I$^-$ migration towards subsequent electrodes [21]. As already mentioned above, degradation due to diffusion of I$^-$ is often a result of the intrinsic decomposition of the perovskite layer and introduction of iodine vacancies in the process. The decomposition of the perovskite layer is, for the most part, detectable in the process of characterization via XRD (from the introduction of $PbI_2$ peaks) [19, 37] and reduction of the absorption spectra, [19, 23] which neither was observed in our case. It is interesting to note that even though Bi et. al have reported a reduction in the absorption spectra even at 15 hours of aging when $FTO/NiMgLiO/CH_3NH_3PbI_3$ was heated at 100 ºC, [23] we observed a different behavior. In our case the absorption spectra remained unchanged at 96 h of aging at 85ºC and $N_2$, similar to what Conings et. al have reported [19]. Some key differences are noteworthy in our case. We have aged films with a different hole transporting layer (which has been shown to affect the perovskite crystallization process in p-i-n structures) [31] with the structure $ITO/PEDOT:PSS/CH_3NH_3PbI_3$ as well as a different aging temperature of 85ºC. In summary, due to the apparent intrinsic stability of $CH_3NH_3PbI_3$ evident from the unchanged XRD and absorption spectra, properly functional devices obtained using buffer layer engineering and aged half device stacks ($ITO/PEDOT:PSS/CH_3NH_3PbI_3$, $ITO/PEDOT:PSS/CH_3NH_3PbI_3/PC[70]BM/AZO$), the change of the perovskite film from dark brown to yellow upon annealing with the Al electrode and the experimental evidence reported in the literature highlighting the correlation between intrinsic perovskite decomposition and $I^-$ diffusion [37] via iodine vacancies, [23] we believe that diffusion of Al into the perovskite layer is the most likely initial cause of degradation in this particular study, similar to what has been previously observed using other common metal electrodes. [21,23,35,36]. Upon perovskite decomposition iodide vacancies can be introduced as reported in the literature [23] promoting $I^-$ diffusion and thus the iodide diffusion mechanism can further contribute towards the degradation of the device which is observable by the Al color change in aged thin-PC[70]BM based devices shown in Figure 5e. Thus, the reported experimental observations suggest that both degradation mechanisms coexist upon thermal aging of devices. The above observations are in agreement with recent literature [23]. We have shown in this paper that thick fullerene-based buffer layers can be used to block diffusion and improve the thermal stability of inverted perovskite PVs.

## 2.4. Reducing the efficiency-stability gap of perovskite photovoltaics

Using a 200 nm thick PC[70]BM ETL we were able to isolate the CH3NH3PbI3 active layer from the metal electrode, a procedure that effectively improves the lifetime performance of device upon heating at 85 $^o$C from 24 hours to over 168 hours. This improved stability however is attained at the cost of strong PCE reduction due to the thick PC[70]BM fullerene-based buffer layer used for the isolation of the perovskite active layer from the top metal electrode. Furthermore, despite the positive effects of the toluene CH3NH3PbI$_3$ washing step in efficiency this processing step can introduce some reliability issues on the reproducibility of the device performance. To reduce the efficiency–stability–reliability gap of perovskite photovoltaics, we have tested the behavior of the proposed diffusion blocking layer to devices that are optimized to work effectively with thick PC[70]BM. These devices were based on a highly reliable perovskite formulation of $CH_3NH_3PbI_{3-x}Cl_x$. Typically, devices based on $CH_3NH_3PbI_{3-x}Cl_x$ produce perovskite films with high roughness (Rms~12.45 nm) compared to the previously used $CH_3NH_3PbI_3$ formulation that produces films with relatively low roughness (Rms~4.21 nm), as shown in the AFM height images in **Figure S3.** The thicker PC[70]BM is therefore necessary to ensure uniform coverage of the rough perovskite films and to improved device reliability.

Using $CH_3NH_3PbI_{3-x}Cl_x$ and thick (200 nm) diffusion based PC70BM blocking layer we have managed to achieve devices with Voc=0.88 V, Jsc=12.99 mA/cm$^2$, FF=62.3 %, PCE=7.11 % (average) and a champion device with Voc=0.90 V, Jsc=15.79 mA/cm$^2$, FF= 66.8 % and PCE= 9.48 % (**Figure S4**). The heat stability of these devices was also tested and compared with the efficient $CH_3NH_3PbI_3$ based devices using optimized for efficiency thin (70 nm) fullerene electron transporting layer (ETL). The lifetime profile of the normalized PV parameters is shown in **Figure 6.** It is observed that the devices which are based on $CH_3NH_3PbI_{3-x}Cl_x$ s with thick PC[70]BM show impressive heat stability for up to 1000h. It has been reported that $CH_3NH_3PbI_{3-x}Cl_x$ exhibits better intrinsic stability over other perovskite formulations such as $CH_3NH_3PbCl_3$ and $CH_3NH_3PbI_3$ [38] it is important to note that based on the experimental conditions used within this paper we haven't observed any major intrinsic degradation of the $CH_3NH_3PbI_3$ from the XRD and UV-Vis. Furthermore, we have shown using buffer layer device engineering experimental methods (please see section 2.3) that using thermally aged $CH_3NH_3PbI_3$ and fresh top electrode yields almost no change to the efficiency parameters of the device compared to a fresh-reference device. Therefore, we believe that the improved stability of the $CH_3NH_3PbI_{3-x}Cl_x$ based devices tested is mainly due to the incorporation of the thick PC[70]BM buffer layer that can be used to eliminate top electrode thermally activated degradation mechanisms of inverted perovskite photovoltaics.

## 3. Conclusion

To conclude we have proposed that the interaction of the Al metal electrode with the perovskite active layer through diffusion mechanisms is the main thermal degradation mechanism for inverted perovskite solar cell. Using a 200 nm thick PC[70]BM ETL we were able to isolate the $CH_3NH_3PbI_3$ active layer from the metal electrode, a procedure that effectively improves the lifetime device performance upon accelerating lifetime heating at 85 $^oC$ from 24 hours to over 168 hours. This improved stability however is attained at the cost of reduced PCE due to the thick PC[70]BM fullerene buffer layer used for the isolation of the $CH_3NH_3PbI_3$ active layer from the top Al metal electrode. To reduce the efficiency–stability gap of perovskite photovoltaics, we have tested the proposed diffusion blocking layer to highly reliable CH3NH3PbI3-xClx based devices that are optimized to work effectively with thick PC[70]BM. We have presented a hysteresis free $CH_3NH_3PbI_{3-x}Cl_x$ inverted perovskite PVs with thick PC[70]BM diffusion blocking layer that exhibit PCE in the range of 10 % and thermal stability for over 1000 hours at 60 $^oC$.

## 4. Experimental section

_$CH_3NH_3PbI_{3-x}Cl_x$ based device fabrication:_ The commercial I201 perovskite precursor ink (Nitrogen Processing) by Ossila Ltd. was used. The precursor was heated at 70$^oC$ inside a $N_2$ glovebox for 2h and then left to cool down to room temperature ITO-patterned glass substrates (sheet resistance 4$\Omega$/sq) were cleaned using an ultrasonic bath for 10m in acetone followed by 10m in isopropanol. The PEDOT: PSS films were prepared using the Al4083 solution from Heraeus. PEDOT: PSS was filtered using 0.22 $\mu$m PVDF filters before coating. The PEDOT: PSS films were coated using spin coating at 6000 rpm for 30s followed by annealing at 150 $^oC$ for 10m. The perovskite films were prepared inside a $N_2$ glovebox using spin coating at 4000 rpm for 30s followed by annealing at 80$^oC$ for 2h. PC[70]BM (50 mg/ml) was then coated using spin coating at 1000 rpm for 30s. The AZO solution used was a commercial solution (N-20X) from Avantama Ltd. AZO films were coated using spin coating at 1000 rpm for 30s. The Al metal was finally thermally evaporated achieving a thickness of 100nm. The devices were then encapsulated using a small glass slide and a UV-curable encapsulation epoxy form Ossila Ltd.

_CH3NH3PbI3_  _based device fabrication:_ The perovskite solution was prepared using a mixture (1:1 M) of $PbI_2$ from Alfa Aesar  and methylammonium iodide (MAI) form GreatCell Solar. The mixture was dissolved in a mixed solvent (7:3) of γ-butyrolactone and dimethyl sulfoxide (DMSO). The solution was stirred at 60 °C for 1h. The preparation of PEDOT: PSS films is identical to the procedure followed for the fabrication of $CH_3NH_3PbI_{3-x}Cl_x$ based devices. The perovskite solution was left to cool at room temperature inside the glovebox followed by filtering using a 0.22μm PVDF filter. The perovskite films were coated using a 3-step spin coating process: 1st step 500 rpm for 5s, 2nd step 1000 rpm for 45 s and 3rd step 5000 rpm for 45 s. During the 3rd step, after the first 20s of the duration of the step, 0.5 ml Toluene is dropped onto the spinning substrate to achieve rapid crystallization of the films. The resulting perovskite films are annealed at 100 °C for 10m. The PC[70]BM  film was coated using a solution of 20 mg/ml and 50mg/ml to achieve fullerene thickness of 70 nm and 200 nm respectively at 1000 rpm for 30s. The AZO and Al coating was identical to the $CH_3NH_3PbI_{3-x}Cl_x$ based formulation.

_Characterization techniques:_ The thickness of the fullerene layers was measured using a Veeco Dektak 150 profilometer. XRD measurements were performed using PANalaytical X's pert pro MPD power diffractometer (40kV, 40 mA) using Cu Kα radiation (λ=1.5418 Å). Optical measurements for the absorption spectra were performed with a Schimadzu UV-2700 UV-Vis spectrophotometer. The AFM measurements were performed using Easy Scan-2 Nanosurf in tapping mode (phase contrast mode) and the roughness was extracted using the Gwydion software. The current density-voltage (J/V) characteristics were characterized with a Botest LIV

Functionality Test System. Both forward and reverse scan directions were measured with 10mV voltage steps and a time step of 40 ms. For illumination, a calibrated Newport Solar simulator equipped with a Xe lamp was used, providing an AM1.5G spectrum at 100mW/cm$^2$ as measured by a certified oriel 91150V calibration cell. A custom-made shadow mask was attached to each device prior to measurements to accurately define the corresponding device area. The photocurrent measurements were performed using a Botest PCT 1 test system. Impedance spectroscopy was performed using an Autolab PGSTAT 302N equipped with FRA32M module. To extract the Nyquist plots, the devices were illuminated using a red LED at 625nm and 100mw/cm$^2$. A small AC perturbation voltage of 10 mV was applied and the current output was measured using a frequency range of 1MHz to 1Hz. The steady-state DC bias was kept at 0V. The C-V measurements for the Mott-Schottky plots were performed under dark using a steady frequency of 5kHz throughout a voltage range of -1V to1V.

## Supporting Information

Supporting Information is available from the Wiley Online Library or from the author.

## Acknowledgements

This project has received funding from the European Research Council (ERC) under the European Union's Horizon 2020 research and innovation programme (grant agreement No 647311). We would like to thank Prof. Nir Tessler for useful discussions

Suppressed decomposition of organometal halide perovskites by impermeable electron-extraction layers in inverted solar cells, *Nat. Commun,* **2017**, *8,* 13938.

**Table of Figures**

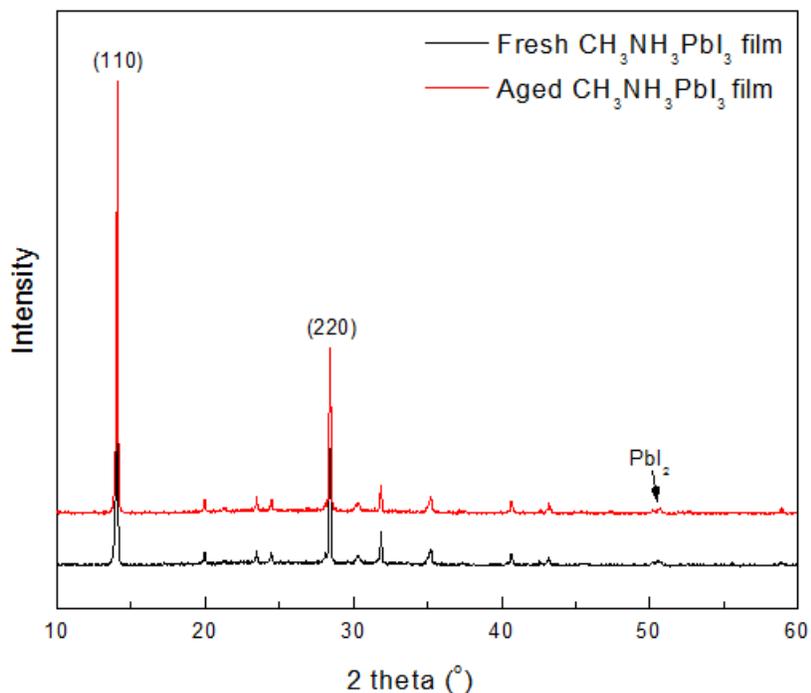

**Figure 1:** XRD patterns of fresh and aged CH$_3$NH$_3$PbI$_3$ films.for 96 h at 85 °C and N$_2$ atmosphere

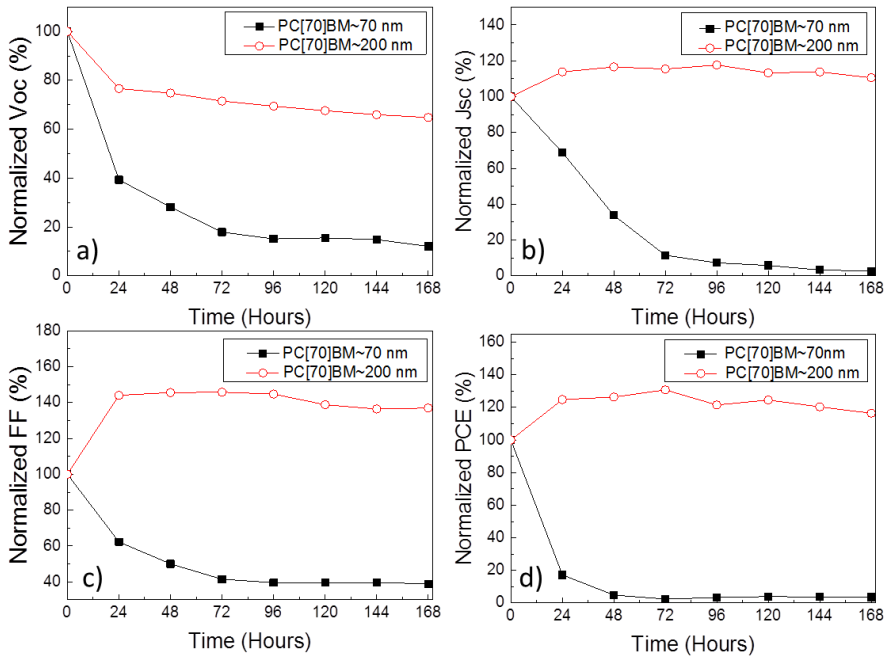

**Figure 2:** Normalized a) Voc, b) Jsc, c) FF and d) PCE of CH$_3$NH$_3$PbI$_3$ based devices over 168 hours of heating at 85 °C using two different fullerene buffer layer thicknesses.

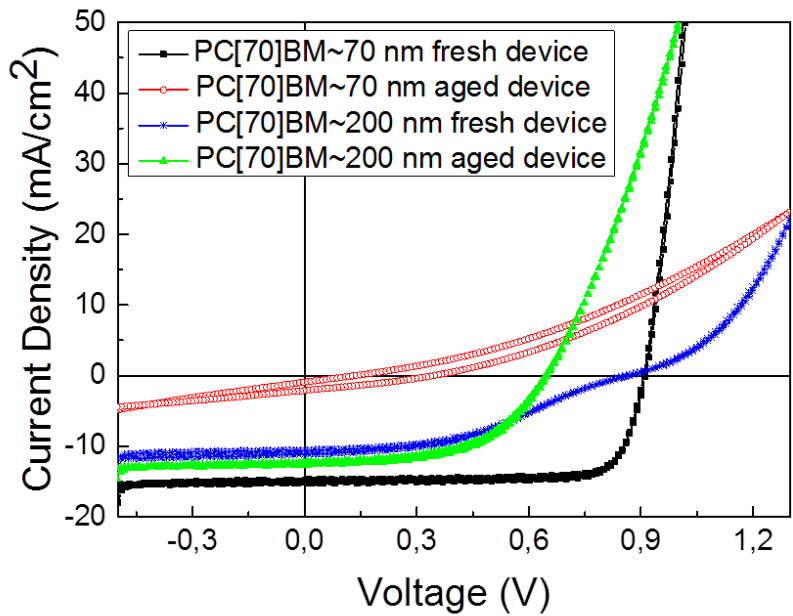

**Figure 3:** Illuminated J/V characteristic for fresh and aged CH$_3$NH$_3$PbI$_3$ based devices.

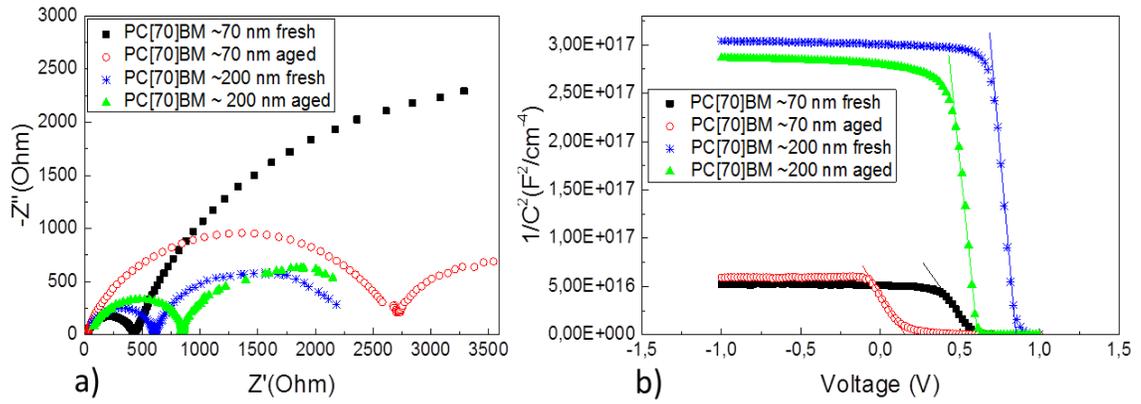

**Figure 4:** a) Nyquist plots, b) Mott-Schottky plots of representative devices

| Device type | Voc (V) | Jsc (mA/cm²) | FF (%) | PCE (%) |
|---|---|---|---|---|
| Fresh reference device | 0.91 | 14.75 | 79.2 | 10.64 |
| Aged ITO/PEDOT:PSS/CH₃NH₃PbI₃ and fresh PC[70]BM/AZO/Al | 0.89 | 15.13 | 67.3 | 9.03 |
| Aged ITO/PEDOT:PSS/CH₃NH₃PbI₃/PC[70]BM/AZO and Fresh Al | 0.84 | 16.43 | 73.2 | 10.10 |

**TABLE 1**. Photovoltaic parameters for fresh reference and buffer layer engineered devices.

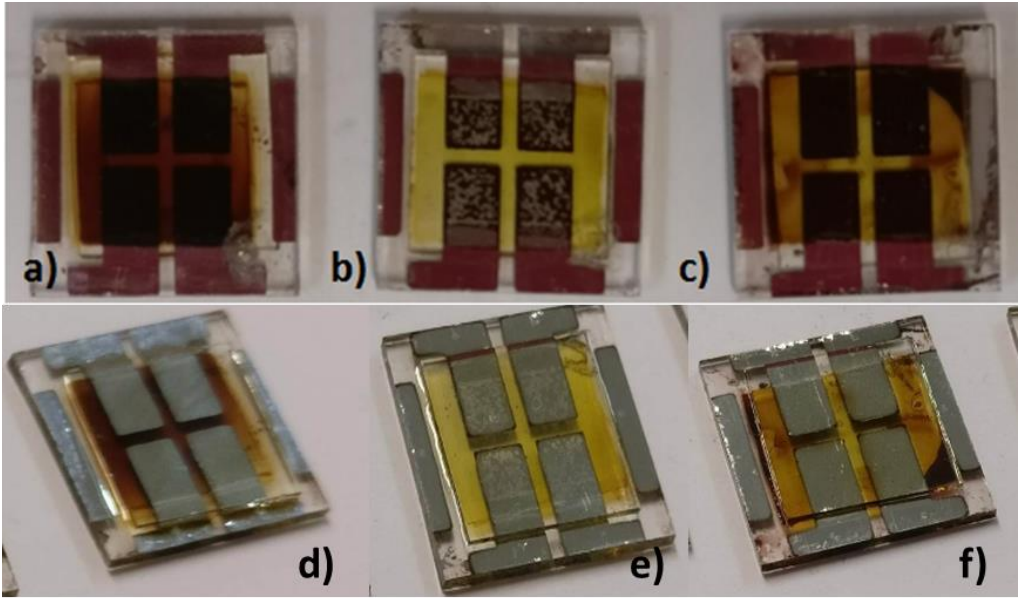

**Figure 5:** a) Bottom side of fresh device stack, b) bottom side of thermally aged device stack of thin-PC[70]BM based devices, c) bottom side of thermally aged devices stack of thick-PC[70]BM based devices, d) top side of fresh device stack, e) top side side of thermally aged device stack of thin-PC[70]BM based devices, f) top side side of thermally aged devices stack of thick-PC[70]BM based devices

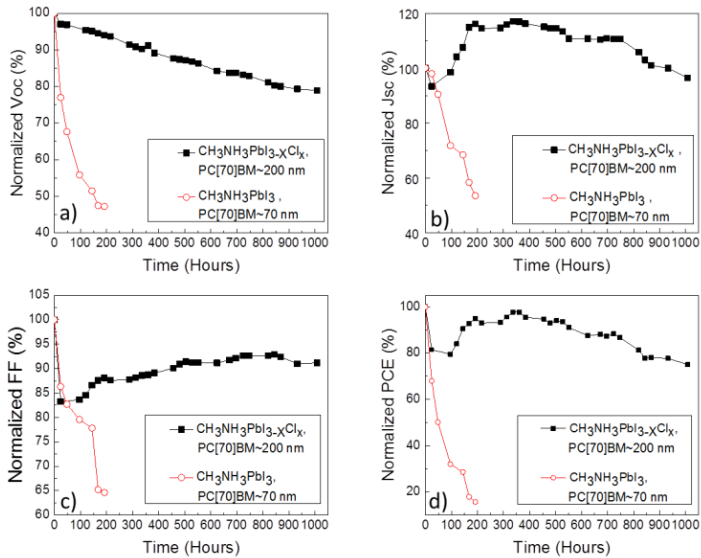

**Figure 6: Normalized a) Voc, b) Jsc, c) FF and d) PCE of CH₃NH₃PbI₃ and CH₃NH₃PbI₃₋ₓClₓ based devices over 1000 hours of heating at 60 ºC and N₂..**

**Table of Contents**

**We present long lived inverted perovskite solar cells (PVSCs) incorporating fullerene-diffusion blocking Layer (DBL).** We showthat the interaction of the top metal electrode (TME) with the perovskite active layer (AL) through diffusion mechanisms is the main thermal degradation mechanism. Using thick fullerene DBL we were able to isolate the AL from the TME and effectively improve heat lifetime performance of inverted PVSCs.

**Keywords:** (perovskites, thermal stability, accelerated heat lifetime, fullerene blocking layers, electrodes)

*Fedros Galatopoulos [1]Ioannis T. Papadas [1], Gerasimos S. Armatas[2] and Stelios A. Choulis [1]\**

# Long Thermal Stability of Inverted Perovskite Photovoltaics

# Incorporating Fullerene-based Diffusion Blocking Layer

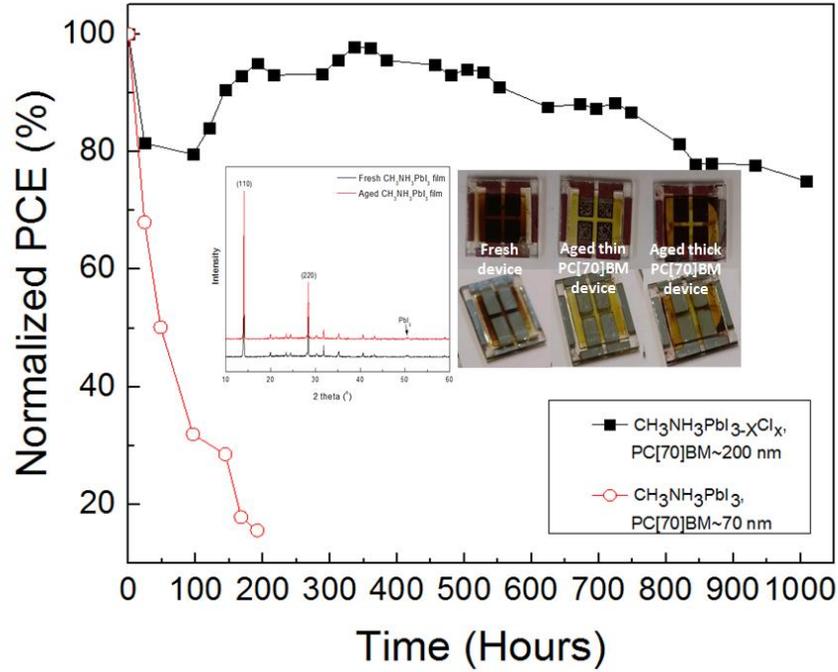

# Supporting Information

## Long Thermal Stability of Inverted Perovskite Photovoltaics

## incorporating Fullerene-based Blocking Layer

*Fedros Galatopoulos [1] Ioannis T. Papadas [1], Gerasimos S. Armatas[2] and Stelios A. Choulis [1]\**

[1] Molecular Electronics and Photonics Research Unit, Department of Mechanical Engineering and Materials Science and Engineering, Cyprus University of Technology, Limassol, 3603 (Cyprus).

[2] Department of Materials Science and Technology, University of Crete, Heraklion 71003, Greece.

*Email: stelios.choulis@cut.ac.cy

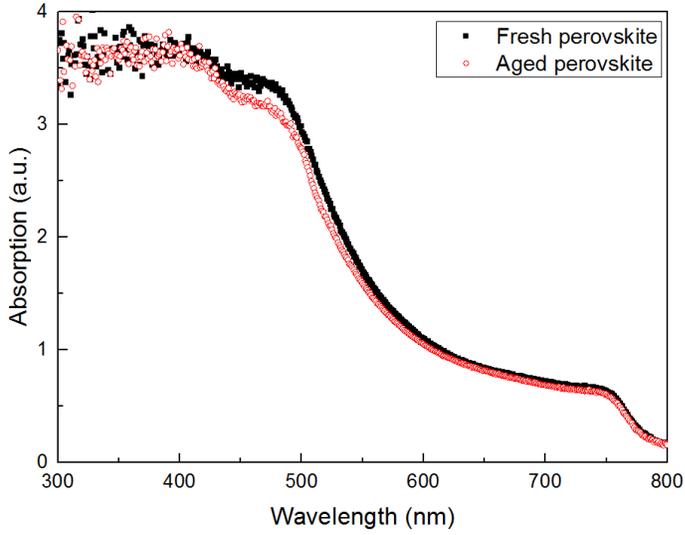

**Figure S1:** Absorption spectra of fresh and aged $CH_3NH_3PbI_3$ films after 96h at 85°C and $N_2$ atmosphere in the dark.

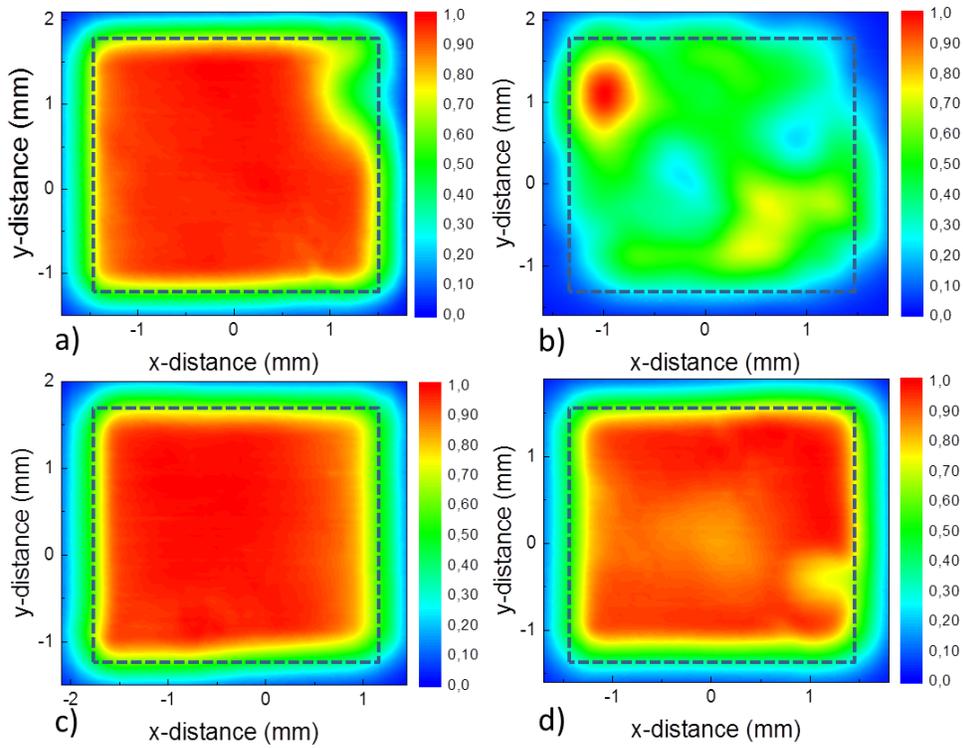

**Figure S2: PCT of a) Thin fullerene fresh, b) thin fullerene aged, c) thick fullerene fresh and d) thick fullerene aged $CH_3NH_3PbI_3$ based device.**

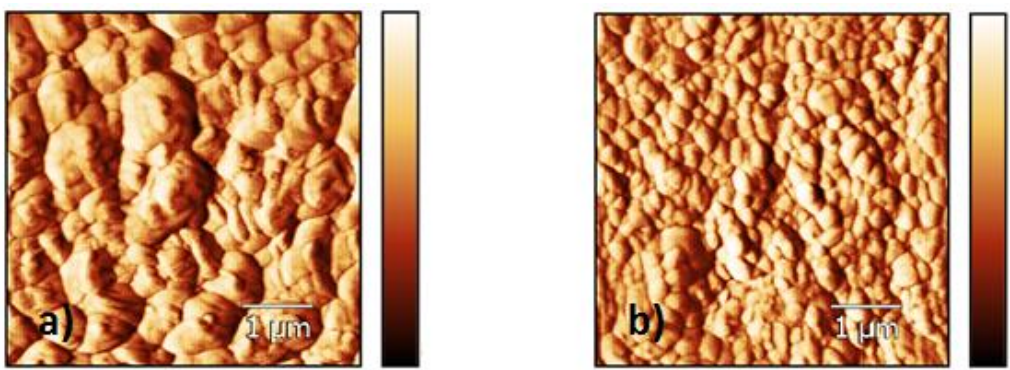

**Figure S3: AFM measurements of a)CH₃NH₃PbI₃₋ₓClₓ and  b) CH₃NH₃PbI₃**

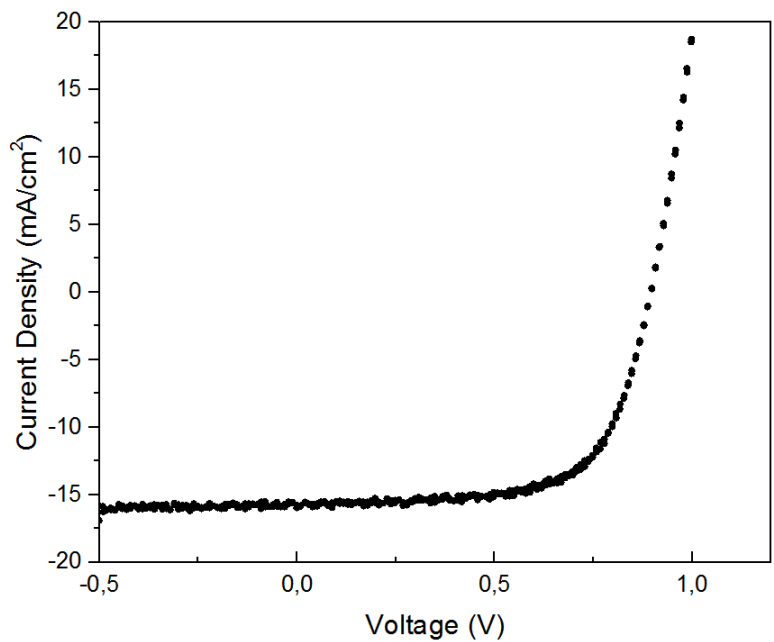

**Figure S4: J/V plot of the champion CH₃NH₃PbI₃₋ₓClₓ based device using thick PC7OBM diffusion blocking layer with PCE in the range of 10 %.**